\documentclass[11pt,3p,review,authoryear]{elsarticle}
\usepackage[utf8]{inputenc}
\usepackage{placeins}
\usepackage{graphicx}
\usepackage{multicol}
\usepackage{latexsym}
\usepackage{amssymb}
\usepackage{amsmath}
\usepackage{amscd}
\usepackage{amsthm}
\usepackage{epsfig}
\usepackage{eucal}
\usepackage{bm}
\usepackage[small,bf]{caption}
\usepackage{lipsum} 
\usepackage[doublespacing]{setspace}				
\usepackage{algorithm}
\usepackage[noend]{algpseudocode}

\journal{International Journal of Forecasting}
\biboptions{longnamesfirst}

\begin{document}

\begin{frontmatter}

\title{A Bayesian Model for Forecasting Hierarchically Structured Time Series}

\author[jn]{Julie Novak \corref{cor}}
\address[jn]{Netflix, California, USA}
%
\author[sm]{Scott McGarvie}
\address[sm]{JP Morgan, London, United Kingdom}
\author[bb]{Beatriz Etchegaray Garcia}
\address[bb]{IBM Thomas J Watson Research Center, New York, USA}
%
\cortext[cor]{Corresponding author. Postal Address: 100 Winchester Cir, Los Gatos, CA, USA, 95032. E-mail address: julien@netflix.com Tel 1-646-346-4238}

\begin{abstract}
An important task for any large-scale organization is to prepare forecasts of key performance metrics. 
Often these organizations are structured in a hierarchical manner and for operational reasons, projections of these metrics may have been obtained independently from one another at each level of the hierarchy by specialists focusing on certain areas within the business. 
There is no guarantee that when combined, these aggregates will be consistent with projections produced directly at other levels of the hierarchy.   
We propose a Bayesian hierarchical method that treats the initial forecasts as observed data which are then combined with prior information and historical predictive accuracy to infer a probability distribution of revised forecasts.  
When used to create point estimates, this method can reflect preferences for increased accuracy at specific levels in the hierarchy.
We present simulated and real data studies to demonstrate when our approach results in improved inferences over alternative methods.  
\end{abstract}

\begin{keyword}
Bayesian methods \sep forecast reconciliation \sep hierarchical time series \sep heterogeneous loss function
\end{keyword}

\end{frontmatter}

\section{Introduction}

Large-scale organizations are often required to provide forecasts of a number of key performance metrics that measure attributes such as their financial health or growth potential \citep[e.g.,][]{hongxia,ibm,fan2008price}.  
These organizations are naturally structured in a hierarchical manner; they conduct business in various locations around the world and are split by divisions that are responsible for distinct products or lines of business. 
Recognizing the importance of accurate forecasts in their decision making, different groups within the organization have often developed their own mechanisms for collecting individual data and creating predictions for their operations in isolation from the rest of the business.   
However, forecasts are often produced independently from one another in the manner outlined above and subsequently collected together to obtain aggregate estimates. 
The independent forecasts will typically be inconsistent with information one has of forecasts provided at other levels of the hierarchy.   

There are a number of different contexts in which the hierarchical structures discussed here might arise in real applications:
corporations such as IBM are structured by line of business and market as discussed above;   
unemployment figures are often disaggregated to state and population demographic levels; 
the S\&P 500, a financial market indicator, is an index based on 500 large American companies which has intermediate levels of aggregation that correspond to industries or sectors.   
Throughout this paper, we will be referring to forecasting revenue for IBM as the motivational example for the method described, however it should be noted that this approach can be applied in many other situations.   

The problem of obtaining forecasts that add up consistently in a hierarchical setting is not new \citep[e.g.,][]{schwarzkopf1988top, shlifer1979aggregation}. 
One commonly used strategy is to forecast an aggregated time series at the highest level which is then allocated to the lower levels based on some estimate of historical proportions, as is done in \citet{schwarzkopf1988top}. 
Alternatively one may choose to produce granular estimates at lowest level in the problem which are then aggregated to obtain higher level forecasts, as in \citet{shlifer1979aggregation}. 
\citet{schwarzkopf1988top} and \citet{shlifer1979aggregation} compare these two approaches that are commonly referred to as ``top-down'' and ``bottom-up'' and provide examples of when one method is preferred over the other.
There are other innovative regression based approaches, however nothing in the literature so far has taken a Bayesian perspective. 

Our proposed approach modifies the forecasts initially provided so that they obey the aggregation constraint imposed by a given hierarchical structure, while at the same time remaining as accurate as possible throughout the levels of interest in the forecasting problem at hand. 
In addition, our approach takes into account the uncertainty across all levels of the hierarchy to obtain the final set of forecasts or the ``revised'' forecasts, as first proposed by \citet{hyndman2011}.
Using a Bayesian framework allows us to incorporate uncertainty across all the levels of the hierarchy via posterior distributions of the revised forecasts. 

We use past accuracy of the forecasts at all levels to determine which data should be given more weight depending on their historical ability to predict. 
We can also incorporate context-specific prior information that a practitioner may have, as this information may not be directly incorporated in the historical data.
For example, in the context of forecasting IBM revenue, senior management may be aware of a divestiture or an acquisition that they would like to integrate into the final set of forecasts. 
This event would affect the projected revenue at all levels of the hierarchy and the ability to incorporate such information is an important element of the model.  

We also include a context-specific heterogeneous loss function to penalize errors in particular nodes differently, depending on the specific problem.  This technique allows one to produce point estimates that reflect the preferences of the model user in a natural way.  For example, in the IBM forecasting revenue problem, the accuracy at the highest level may have more influence than that of the lower levels, as this value corresponds to the total revenue of the company to be used by the CEO to make organizational decisions. 
We would pick a point estimate from the posterior that maximizes the utility (which in this example is the top level).
On the other hand, if the company is trying to decide whether to open a new office, it would be more important to have accurate forecasts at a lower level of the hierarchy. 

To illustrate the ability of the proposed approach to achieve improvement in the forecasts, we compare our method to that of several competitor methods which we discuss in more detail below.
We include two sets of studies: one with simulated and one with IBM revenue data.
By simulating a range of datasets (and their corresponding forecasts), we evaluate the competing versus our own methods under different scenarios. The metrics we use to compare are aggregate consistency and accuracy of the initial set of independent forecasts.

The remainder of this work is structured as follows.
In Section \ref{current}, we review existing methods that are used to obtain aggregate consistent forecasts from a set of initial estimates.
In Section \ref{estimation}, we propose a Bayesian approach for modeling hierarchically structured time series data. 
Our model takes into account the disaggregated forecasts for each individual series, their historical accuracy, and any prior information provided by the practitioner, all the while maintaining the aggregate consistency required by the structure of the hierarchy. 
In Section \ref{hetero}, we motivate the need for a heterogeneous loss function, and explain how to use it to obtain revised forecasts for the entire hierarchy from the full posterior distribution.
In Section \ref{sims}, we simulate a range of datasets (and their corresponding forecasts) and compare our method to four alternative approaches. In Section \ref{realdata}, we apply our Bayesian approach to IBM revenue data and demonstrate consistent results with the previous section. 
Finally, in Section \ref{conc}, we summarize our contribution and discuss future directions for this work. 

\section{Existing Methods in Hierarchical Forecasting} \label{current}
We begin with a simple example of a hierarchically structured problem which is shown schematically in Figure 1.
Here each node represents a time series of monthly revenue for a different geographic market and product offering combination of a single company.
In the case of a large organization such as IBM, the hierarchical structure may be much larger. 
The problem at hand is to forecast revenue for this company using historical observations of revenue.
In order to solve our stated problem we must forecast monthly revenue in each of the cells in Figure 1.

\begin{figure}[h]
\begin{center}
\includegraphics[width=4.5in]{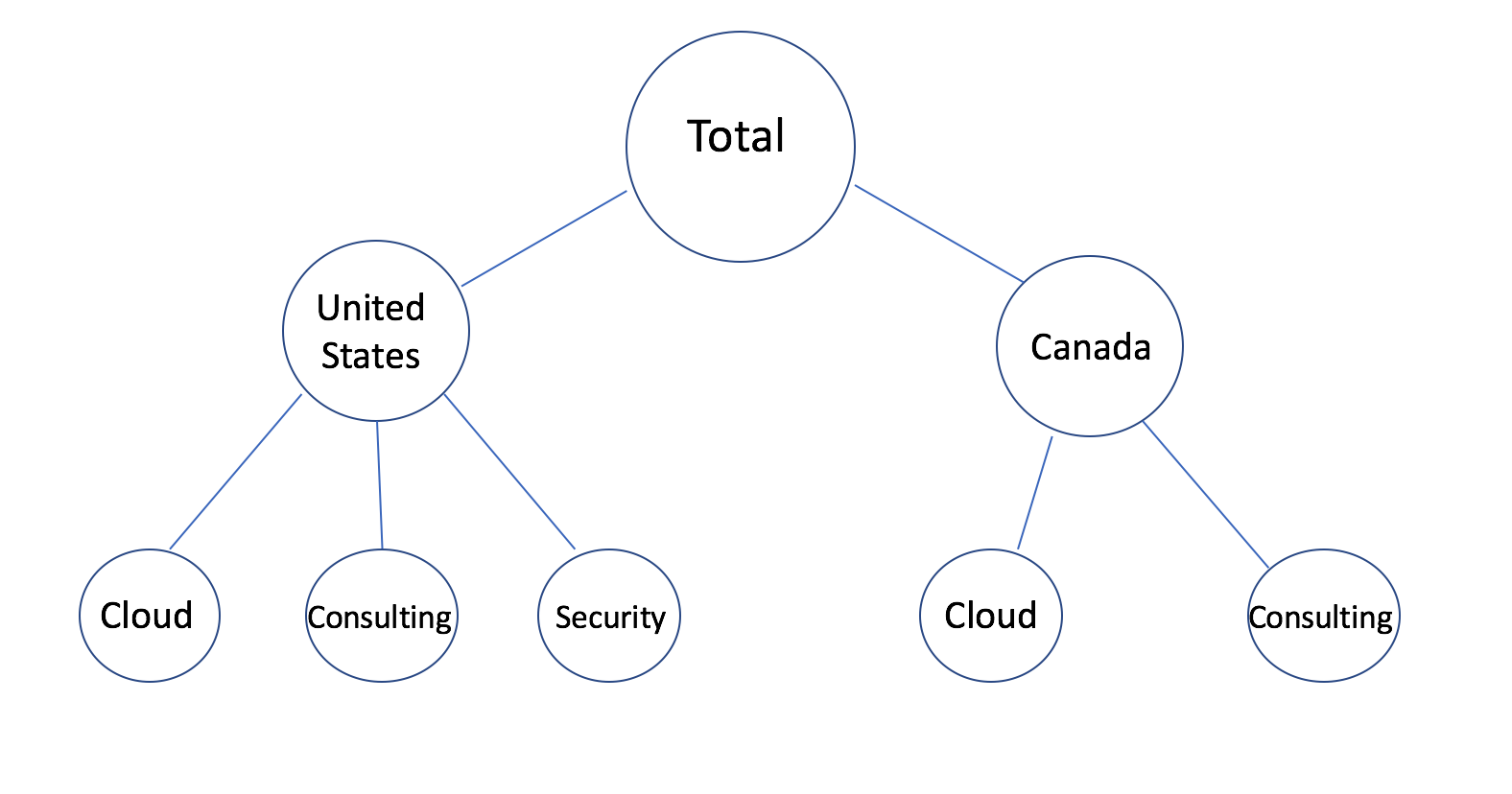}
\end{center}
\vspace{-1cm}\caption{A simple example of a hierarchical organizational structure divided by market and then by division.  \label{fig:hierarchy}}
\end{figure}
\FloatBarrier

The top level, referred to as `Total' in Figure \ref{fig:hierarchy}, would be the total revenue of the company.
It is obtained as the sum of the revenue for the different countries in the second level.
Similarly, for each country, the revenue is the sum of the revenues of the product offerings for that country.
We denote by $K$ the number of levels in a hierarchy.
In the case of Figure \ref{fig:hierarchy}, $K=3$.

Let $\pmb{S}$ be the ``summing'' (or aggregation) matrix that defines all the nodes in the hierarchy in terms of the lowest level, as proposed by \citet{hyndman2011}. 
For the hierarchy in the example given in Figure \ref{fig:hierarchy}, this would be

\begin{equation}
{\small
\underbrace{\begin{bmatrix}
y_{1} \\ y_{21} \\ y_{22} \\ y_{311} \\ y_{312} \\ y_{313} \\ y_{321} \\ y_{322}
\end{bmatrix}}_{\pmb{Y}}
=
\underbrace{\left[\begin{matrix}
1 & 1 & 1 & 1 & 1  \\
1 & 1 & 1 & 0 & 0  \\
0 & 0 & 0 & 1 & 1  \\
1 & 0 & 0 & 0 & 0   \\
0 & 1 & 0 & 0 & 0   \\
0 & 0 & 1 & 0 & 0  \\
0 & 0 & 0 & 1 & 0   \\
0 & 0 & 0 & 0 & 1  \\
\end{matrix} \right] }_{\pmb{S}}
\times
\underbrace{\left [\begin{matrix} y_{311} \\ y_{312} \\ y_{313} \\ y_{321} \\ y_{322}  
\end{matrix} \right] }_{\pmb{Y_K}}
}
\label{eqn:Smatrix}
\end{equation} 
where $\pmb{Y}$ is the vector of all the observations in the hierarchy at a particular time $t$, $\pmb{S}$ is the summing matrix as described above, and $\pmb{Y}_K$ is the vector of all the observations in the lowest level of the hierarchy at time $t$.

The summing matrix $\pmb{S}$ in Equation \ref{eqn:Smatrix} is of dimension $(m \times m_K)$ where $m$ corresponds to the total number of nodes in the hierarchy and $m_k$ corresponds to the number of nodes at the lowest level.  In the current example we therefore have $m = 8$ and $m_K = 5$.  We will also use the notation $y_{321}$, where the number of indices refers to the level of the hierarchy we are referring to. The first number specifies the level, the second specifies the node at that level, and the third specifies the node at the following level. For example, $y_{321}$ corresponds to the observation  at the third level, second node of the second level, and first node of that second node (at the third level). In Figure \ref{fig:hierarchy}, $y_{321}$ corresponds to the observation in the `Canada/Cloud' node.

The problem of forecasting in a hierarchical structure consists of producing estimates for each of the series in the hierarchy that are also aggregate consistent.
We can think of $\pmb{Y}_{1:(t-1)}$ as the observations and $\hat{\pmb{Y}}_t$ as the forecasts generated from some established business process.
These may be independent forecasts generated for each of the time series in the hierarchy and do not necessarily satisfy Equation \ref{eqn:Smatrix} as they have been produced without knowledge of the forecasts for other nodes.
Therefore, it is necessary to ``revise'' or consolidate these forecasts in a way that they become aggregate consistent. 


One commonly used strategy is to focus only on the top-most forecast of the hierarchy and then to disaggregate down to the lower levels based on some specified model. 
For example, using the historical proportions of bottom level series relative to upper level ones.
This method is typically referred to as ``top-down'' forecasting.
One particular problem with this approach is that dynamics that exist at higher levels of the hierarchy may not exist in the lower ones.
The top-down approach will produce revised forecasts that will be biased towards the top level dynamics.
Alternatively, in the ``bottom-up'' approach, one starts by forecasting the series at the lowest level and then adds these up to obtain upper level forecasts.
A problem with the bottom-up approach however is that the bottom level time series may be noisy and volatile and thus hard to forecast.
A ``middle-out'' approach combines the two previously described methods: forecasts are first obtained at some intermediate level of the hierarchy and then the bottom-up method is used to forecast the upper levels while the top-down approach is used for the lower levels.
The biggest drawback in any of these approaches is that they do not use all the available information.
Specifically, the covariance structure of the hierarchy is not incorporated into the model.   

\citet{hyndman2011} observed that the three methods described can be written as a linear mapping from a set of initial forecasts (that are not aggregate consistent) to reconciled forecasts (that are).
In an effort to optimally produce hierarchical forecasts, the authors proposed a general hierarchical forecasting framework that takes independent forecasts at all levels of disaggregation (i.e., the ``base'' forecasts) and reconciled them using a regression model. \citet{tourism} demonstrate the usefulness of forecast reconciliation methods in the context of tourist demand. \citet{hyndman2016} extend their previous work by using weighted least squares and thus taking into account the variability of the base forecast errors. 

None of these approaches produce probabilistic forecasts, which are needed to quantify uncertainty for all the nodes in the hierarchy, as emphasized in \citet{berrocal2010probabilistic}. 
One proposal for addressing this is a multi-step algorithm was made by \citet{taieb2017coherent} by using copulas and LASSO. 
However, none of the previous approaches have taken a Bayesian perspective, the benefits of which will be described below.

\section{A Bayesian Model for Hierarchically Structured Time Series Data} \label{bayesian}
We begin with a set of $m$ independent forecasts $\hat{\pmb{Y}}_t$ (for a future time point $t$).
These forecasts $\hat{\pmb{Y}}_t$ do not necessarily add up according to the structure of the hierarchy. 
In fact, it is highly unlikely that forecasts obtained in such a manner will be aggregate consistent. 
We refer to $\pmb{Y}_t$ as the truth (which is always aggregate consistent), whereas $\hat{\pmb{Y}}_t$ are the forecasts acquired using the practitioner's choice of forecasting method. 
Now, as proposed by \citet{hyndman2011}, we can think of Figure \ref{fig:hierarchy}, the full hierarchical structure, as a linear model,

\begin{equation}
    \hat{\pmb{Y}}_t\sim N(\pmb{S} \pmb{\beta}_t, \pmb{\Omega}_t)
\end{equation}
where $\pmb{\beta}_t$,  of dimension $m_K \times 1$, are the unknown means of the bottom level $K$ nodes for a future time point $t$,
and $\pmb{\Omega}_t$ is the covariance matrix that represents the deviations of the supplied forecasts around the aggregate consistent ones (the mean value of $\pmb{S} \pmb{\beta}_t$).
 
As a result, the initial forecasts are noisy observations of the expected value of the series.
An assumption of this model is that the initial forecasts are unbiased estimates, so that the noise has expected value of zero.
The unknown $\pmb{\beta}_t$ can be estimated using the generalized least-squares procedure,

\begin{equation} \label{eq:gls}
\min\limits_{\beta_t} (\hat{\pmb{Y}}_t - \pmb{S} \pmb{\beta}_t)^T \pmb{\Omega}_t^{-1} (\hat{\pmb{Y}}_t - \pmb{S} \pmb{\beta}_t).
\end{equation}

A problem with Equation \ref{eq:gls} is the difficulty of estimating $\pmb{\Omega}_t$. \citet{hyndman2011} make an assumption that leads to using $\pmb{\Omega}_t$ = $\pmb{I}$ as the identity matrix. This assumption works well in some cases \citep[see][]{tourism}. Moreover, the covariance matrix of the aggregation errors is non-identifiable, as shown by  \citet{wickramasuriya2015forecasting}. Strategies for estimating $\pmb{\Omega}_t$ for reconciliation have been made by \citet{hyndman2016} and \citet{wickramasuriya2015forecasting} by using sparse methods to reduce computational complexity in large hierarchies. 

There are downsides to using simplistic methods such as bottom-up, top-down, and middle-out and there are difficulties in applying the generalized least-squares framework to the hierarchical time series forecasting problem. In the latter case, the reconciliation step ties the aggregate consistency requirement to sharing information across the hierarchy.
We consider a two-step approach for reconciliation of the initial forecasts. 
First we optimally learn the distribution of base or bottom-level forecasts using the information across all levels of the hierarchy with a Bayesian approach, and then we obtain aggregate consistent point forecasts.

\subsection{Model Estimation} \label{estimation}
We initially make the assumption that $\pmb{\Omega}_t$ is diagonal. This assumption is often appropriate and represents a structure where each of the base forecasts are taken to be independent of one another.
We then extend to the case where $\pmb{\Omega}_t$ is block diagonal. 
By imposing a block diagonal structure, we allow for interdependence across the series which have a parent/child node relationship.


We decompose the covariance matrix, $\pmb{\Omega}_t $, as the product of two terms: 
\[
\pmb{\Omega}_t = \pmb{Q}_t \sigma^2_t 
\]
where the matrix $\pmb{\Omega}_t $ is known up to a scalar factor. 
The $\sigma^2_t$ is the component of the covariance matrix of the aggregation errors that is not associated with the forecasting procedure used to obtain the initial forecasts $\hat{\pmb{Y}}_t$.
The variance component $\sigma_t^2$ is unknown \citep[see][for details]{bda2}.
The idea behind this specific decomposition is to separate the covariance components into those that can be attributed to the forecasting method used and those that cannot. 
The matrix $\pmb{Q}_t$ will consist of the forecast errors for all the nodes. 

We first compute a diagonal $\pmb{Q}_t$ based on the historical accuracy of the $m$ individual forecasts for each node of the hierarchy.
We allow each diagonal entry of $\pmb{Q}_t$ to be the mean squared error associated with a specific forecasting method used for that level.

\begin{equation}
\pmb{Q}_t = \begin{bmatrix}
\gamma_{1,t} & 0 & \ldots & 0\\
 0 &\gamma_{2,t} & \ldots & 0  \\
\vdots & \vdots & \ddots & \vdots \\
 0 & 0 & \ldots & \gamma_{m,t}  
\end{bmatrix}
\end{equation}
We recommend using historical accuracies for the $m$ individual forecasting methods to create the $\pmb{Q}_t$ matrix. 
For example, for the historic accuracy at each node of the hierarchy, we use the average of the last 20 percent of the time series' mean squared errors. 
Let $R$ represent the size of the holdout set used to evaluate historical accuracy.
\begin{algorithm}
\begin{algorithmic}[1]
\For{$i=1,\ldots, m$}
\For{$r=1,\ldots, R$}
\State \texttt{holdout last $r$ values} \Comment Apply procedure for $R$ holdout steps
\State \texttt{forecast one step ahead}
\State \texttt{compute $squared\_error_{r}$} $=(forecast-holdout )^2$
\EndFor
\State \textbf{return} $\gamma_{it} = \sum_{r=1}^{R} squared\_error_{r}/R$\Comment{This is the mean squared error for each node}
\EndFor
\end{algorithmic}
\end{algorithm}
\FloatBarrier

\begin{figure}[h]
\begin{center}
\includegraphics[width=4.5in]{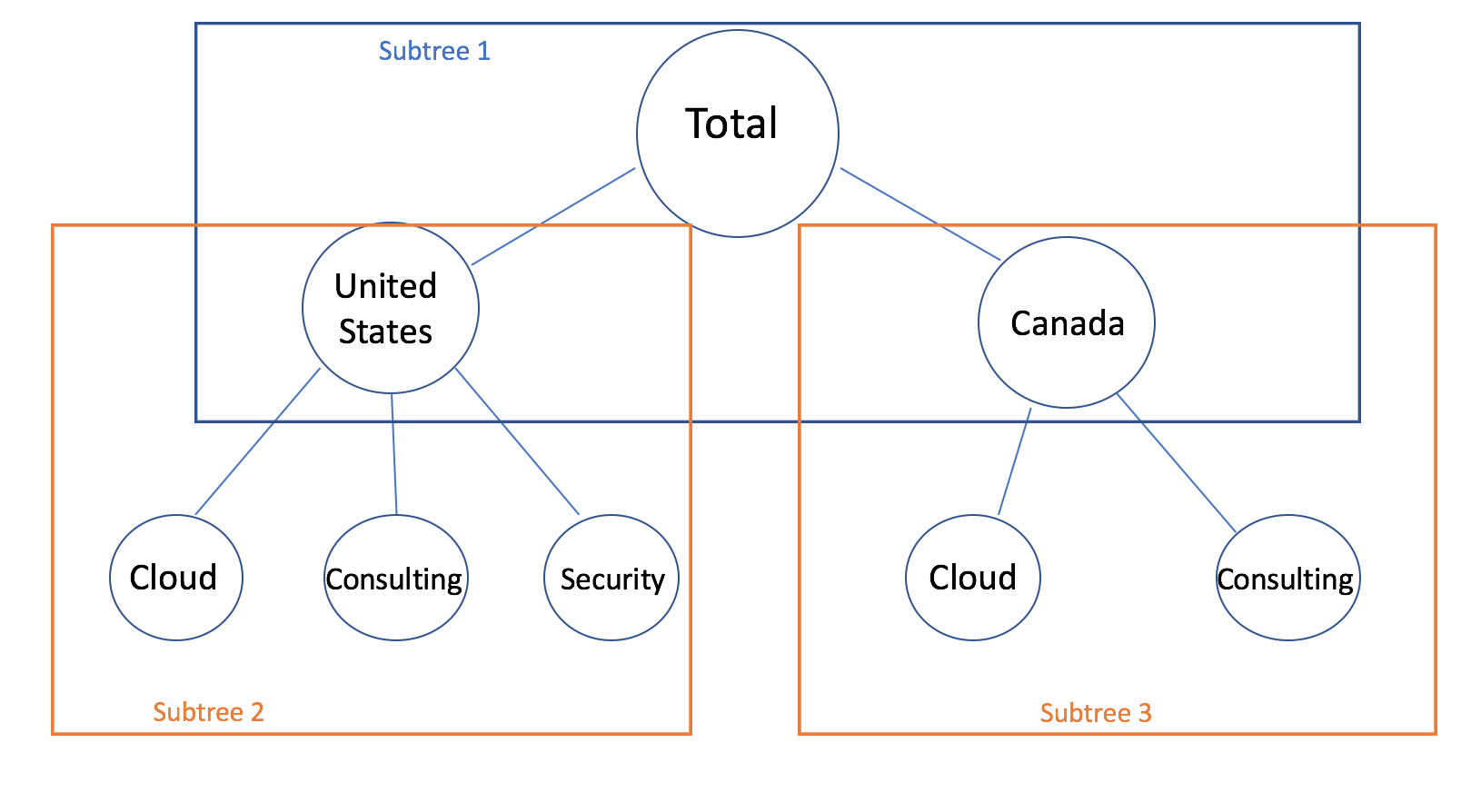}
\end{center}
\vspace{-1cm}\caption{Example of the subtree would be divided in our example.  
\label{fig:hierarchy_tree}}
\end{figure}

If we do not assume the diagonal structure, in Figure \ref{fig:hierarchy_tree} we provide an example of how to split a dataset into subtrees to structure the matrix $\pmb{Q}_t$. 

\FloatBarrier

Each subtree consists of a parent node and all its children. To compute the block diagonal version of this matrix, we first compute the diagonal elements exactly as was done in the first $\pmb{Q}_t$. 
Then the off diagonal would be equal to the relative accuracy of one child node compared to all the other child nodes of that same parent. In other words, we want the more accurate child to have more weight with its parent node than the other children.
The off diagonal entries will be the proportion of the accuracy of the parent node that is attributed to the accuracy of the child node.

In Figure \ref{fig:hierarchy_tree}, the example for the $Q_{\text{US; US/Cloud},t}$ entry between `US' and `US/Cloud' would be

\begin{equation}
Q_{\text{US; US/Cloud},t} = \frac{\gamma_{\text{US/Cloud},t}}{\gamma_{\text{US/Cloud},t}+\gamma_{\text{US/Consulting},t}+\gamma_{\text{US/Security}},t} \gamma_{\text{US},t}
\label{eq:offdiag}
\end{equation}
The higher the accuracy of the `US/Cloud' node ($\gamma_{\text{US/Cloud},t}$) the more we would want it to correlate with the `US' parent node (as compared to the other child nodes, `US/Consulting' and `US/Security'). 

To ensure that $\pmb{Q}_t$ is positive definite, we need to set every second full level of subtrees to zero.
For example, in Figure \ref{fig:hierarchy_tree}, we either set subtree 1 to zero or both subtrees 2 and 3 to zero.



We place a noninformative prior distribution on $(\pmb{\beta}_t,\log \sigma_t)$ \citep[][]{bda2}, or equivalently
\begin{equation}
    \pi(\pmb{\beta}_t, \sigma_t^2) \propto \sigma_t^{-2}
\end{equation}


To sample draws of the posterior distribution of the parameters of interest, $(\pmb{\beta}_t,\sigma^2_t)$, we first solve for $\pmb{Q}_t^{-1}$. 
When the tree is large, inverting $\pmb{Q}_t$ will be also be straightforward, since we set up $\pmb{Q}_t$ to be either a diagonal or block diagonal matrix. 

We estimate the joint posterior of our model parameters $\pmb{\beta}_t$ and $\sigma_t^2$ using a Gibbs sampler which is a Markov chain Monte Carlo method \citep[originally proposed by][]{geman1984stochastic, rubin1986multiple}.
Given the way we set up the prior distribution, we draw $\pmb{\beta}_t$ from a Gaussian distribution,
\begin{equation}
   \pmb{ \beta}_t| \sigma^2_t,\hat{\pmb{Y}}_t \sim N(\pmb{\hat{\beta}}_t,\pmb{V}_{\beta_t} \sigma^2_t)
       \label{eq:beta}
\end{equation}
and we draw the scalar variance parameter $\sigma_t^2$ from an scaled inverse-$\chi^2$ distribution,
\begin{equation}
    \sigma_t^2|\hat{\pmb{Y}}_t \sim Inv-\chi^2 (m-m_K,s^2)
       \label{eq:sigsq}
\end{equation}
where this means replacing the mean and variance in equation (\ref{eq:beta}) by
\begin{equation}
    \pmb{\hat{\beta}}_t = (\pmb{S}^T \pmb{Q}^{-1}_t \pmb{S})^{-1} \pmb{S}^T \pmb{Q}^{-1}_t \hat{\pmb{Y}}_t
\end{equation}
\begin{equation}
    \pmb{V}_{\beta_t} = (\pmb{S}^T \pmb{Q}^{-1}_t\pmb{S})^{-1}
\end{equation}
and replacing the scale parameter in equation (\ref{eq:sigsq}) by
\begin{equation}
    s^2 = \frac{1}{m-m_K}(\hat{\pmb{Y}}_t-\pmb{S}\pmb{\hat{\beta}}_t)^T\pmb{Q}^{-1}_t(\hat{\pmb{Y}}_t-\pmb{S}
    \pmb{\hat{\beta}}_t)
\end{equation}

At this point, we have obtained the joint posterior distributions for the lowest level series, $p_{\pmb{\beta}_t}$. 
Note that we used the information across \textit{all} the levels of the hierarchy in order to do so through the transformation $\pmb{S} p_{\pmb{\hat{\beta}}_t}$ and the $\pmb{Q}^{-1}_t$ matrix.



Now we have the posterior forecast distributions for the bottom level nodes of the hierarchy.  The next step consists of obtaining a set of point forecasts that are aggregate consistent. 

\subsection{Minimize the Expected Loss} \label{hetero}


To find a set of \emph{point} forecasts from the posterior that add up properly, we maximize the utility defined by the practitioner.
%
%
This requires incorporating a heterogeneous loss function. We have covered the practical considerations of implementing an MCMC; convergence of the chains, burn-in period, and thinning to avoid correlation of sequential samples. These chains are used to create the point estimates.

 Let $\pmb{\tilde{Y}}^{\star}_t$ be the set of revised point forecasts  that minimize the expected heterogeneous loss function (provided by the practitioner for the context of the problem). 
If there is no preferences for loss function, we recommend using a squared error loss with the same weight on each node, or equal weights across each level of the hierarchy. 
We are looking for a $\pmb{\tilde{Y}}^{\star}_t$ such that 

\begin{equation}
    \pmb{\tilde{Y}}^{\star}_t =  \underset{\pmb{\beta}_t \in p_{\hat{\pmb{\beta}}_t}}{\text{argmin }} \mathbb{E}[\mathcal{L} (\pmb{S} \pmb{\beta}_t,p_{\pmb{\tilde{Y}}_t})]
\end{equation}
where $\pmb{\beta}_t$ are draws from the posterior of $\pmb{\hat{\beta}}_t$, $\mathcal{L}$ is a loss specified by the practitioner, and $p_{\pmb{\tilde{Y}}_t}$ is the distribution of forecasts for all levels.
The posterior distribution $p_{\pmb{\tilde{Y}}_t}$ is obtained by mapping the lower level distributions $p_{\hat{\pmb{\beta}}_t}$ to all levels of the hierarchy using the summation matrix $\pmb{S}$.

The optimal $\pmb{\tilde{Y}}^{\star}_t$ is aggregate consistent with the minimal expected loss with respect to all the posterior draws of the sampler. 
By finding a $\pmb{\tilde{Y}}^{\star}_t$ in such a manner, we are simultaneously incorporating the uncertainty from the posterior distribution of all the cells of the hierarchy while penalizing cells differently based on the heterogeneous loss specified by the practitioner. 


Defining a loss function that is in line with the business context will be an important step when using our methodology in practice. 
In a real business context, higher accuracy would be required at some levels more than at others. 
There would be larger repercussions for the business if the estimates are incorrect at some levels of the hierarchy and we should penalize some errors more than others depending on the context of the problem.
Specifying what nodes require higher levels of accuracy is a critical step in the updating scheme of our reconciliation method.

Examples of choices for the loss function are:
\begin{itemize}
	\item A weighted squared error loss with a highest weight on the top level, if accuracy at the top level is most important.    
	\item Asymmetric loss.  This would be appropriate if the consequences in under and over estimating are different. For example, if the business is deciding on whether to construct a new facility.  
\end{itemize}


\section{Simulation Results} \label{sims}
\begin{figure}[h]
\begin{center}
\includegraphics[width=4.5in]{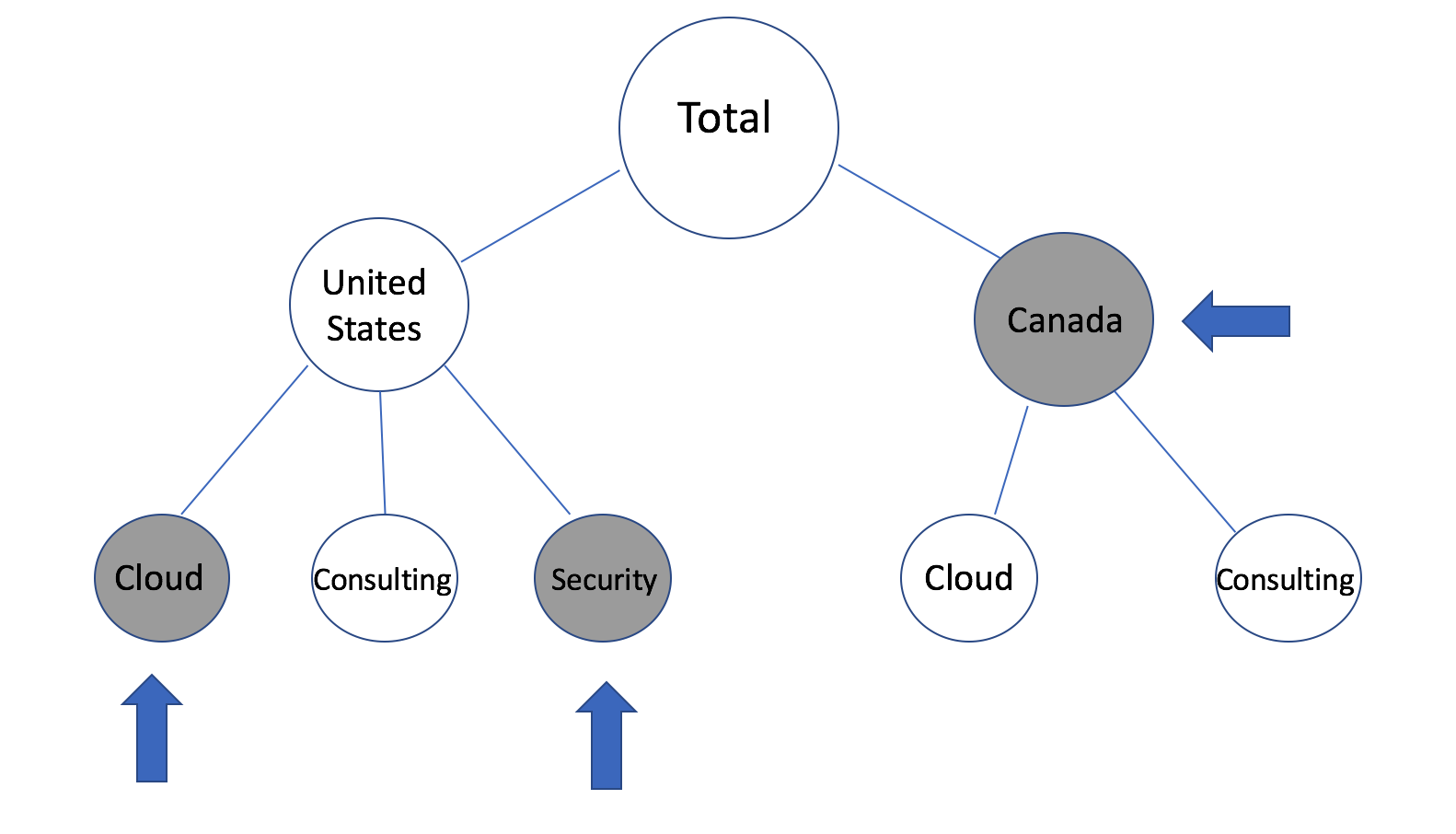}
\end{center}
\vspace{-1cm}\caption{In our simulation setting, certain nodes of the hierarchy, specifically the ones that are shaded in grey, are easier to forecast than others.  
\label{fig:hierarchy2}}
\end{figure}

We consider several simulated data settings to explore how our proposed Bayesian method performs relative to competing methods. 
The simulated data will be set up as shown in Figure \ref{fig:hierarchy2}. For every node in the lowest level of the tree we simulate time series data. To obtain forecasts, we find the best ARIMA model that fits the simulated data using the default settings of the auto.arima function of the forecast package in R\citep[][]{forecastR, khandakar2008automatic}. The example could be considered a small version of the IBM market and division hierarchy. 
A desirable outcome is one where a model's revised forecast in an accurate node does not deviate from the base forecast.
To test this property, we set up the simulated data as follows.
The nodes of the tree that are shaded in gray (`Canada', `US/Cloud', and `US/Security') have more accurate forecasts than the remaining nodes under certain simulation settings.

\FloatBarrier

We compare our results to four competing methods: bottom-up (BU), top-down (TD), the ordinary least squares approach of \citet{hyndman2011} (OLS), and the weighted least squared approach of \citet{hyndman2016} (WLS).  
We refer to our method as Bayesian Reconciliation (BR).
In the top-down method, we predict the top level and then split up the prediction into the lower levels according to historical proportions.
In addition, we compare two settings of our method: using a diagonal and a block diagonal for the $\pmb{Q}_t$ matrix as described in Section \ref{bayesian} (which we call $\pmb{Q1}$ and $\pmb{Q2}$, respectively).



We first define two concepts which we will use throughout the remainder of the paper. 
Forecastability of a node is the metric used to quantify the magnitude of the forecast error.
If $\hat{Y}_{t,i}$, the given forecast for node $i$, is very far from the truth, $Y_{t,i}$, then the forecastability for that node is high, in other words it is hard to independently forecast that node with the information we are provided.
However if $\hat{Y}_{t,i}$, the given forecast for node $i$, is very close from the truth, $Y_{t,i}$, we consider that low forecastability, as it is easy to forecast that node given the information we have. The forecastability for node $i$ is defined as,

\begin{equation}
\text{forecastability}(Y_{t,i}) = \left|\frac{\hat{Y}_{t,i} - Y_{t,i}}{Y_{t,i}}\right|.
\end{equation}

The second metric quantifies how aggregate consistent the base forecasts of the children are with respect to the parent. 
If the nodes add up properly, then the methods would not need to adjust the base forecasts from what was originally provided, and we call that low aggregate consistency. 
If the nodes do not add up at all, the methods would need to adjust them and deviate substantially from the base forecast. We call this high aggregate consistency. The aggregate consistency for node $i$ is defined as,

\begin{equation}
\text{aggregate consistency}(Y_{t,i}) = \left|\sum_{j}\hat{Y}_{t,i+1,j} - \hat{Y}_{t,i}\right|,
\end{equation}
where $j$ indexes the children of node $i$.

We generate 2,000 datasets in which we systematically vary the length of the time series and the magnitude of the base forecasting error of the series at the lowest level. 
We split the data into ``easy'' to reconcile (datasets that fall in into the intersection of low aggregate consistency and low forecastability) and ``hard'' to reconcile (datasets that fall into the intersection of high aggregate consistency and high forecastability). 
For every simulated dataset, we compute a measure of aggregate consistency and forecastability. To define high and low values of these metrics we use cutoffs based on quantiles. 

We say a dataset is highly forecastable if it falls into the highest quantile, i.e., 
\begin{equation}
|\hat{Y}_{\text{Canada}, t} - Y_{\text{Canada}, t} | < \text{Quantile}1,
\end{equation} 
and a low forecastable dataset will be one that satisfies
\begin{equation}
|\hat{Y}_{\text{Canada}, t} - Y_{\text{Canada}, t} | > \text{Quantile}3.
\end{equation}
In other words, if the base forecast $\hat{Y}_{\text{Canada}, t}$ is further from the truth than 75 percent of datasets then it has low forecastability and if it is closer to the truth than 75 percent of datasets it has high forecastability. 

Similarly, we say that a dataset is highly aggregate inconsistent if it satisfies
\begin{equation}
(\hat{Y}_{\text{Canada/Cloud},t} + \hat{Y}_{\text{Canada/Consulting},t}) - \hat{Y}_{\text{Canada},t} > \text{Quantile}3,
\end{equation}
and an aggregate consistent dataset in our setting would be  
\begin{equation}
(\hat{Y}_{\text{Canada/Cloud},t} + \hat{Y}_{\text{Canada/Consulting},t}) - \hat{Y}_{\text{Canada},t} < \text{Quantile}1.
\end{equation}
In other words, we are looking for high and low deviation of the sum of children to parent nodes in the `Canada' subtree.

Since there is no standard method to assess forecasting accuracy in a tree, we focus on different subsets of the tree to assess overall performance. 
In practice, accuracy will be required at different nodes depending on the problem. We compare methods based on two different settings. The final measure of performance will be a weighted average across the entire tree. 

In Setting 1, we ignore the extremely noisy nodes that have highly forecastability. 
We place equal weight of $\frac{1}{6}$ on all the nodes of our simulated hierarchy except for `Canada/Cloud' and `Canada/Consulting', which each get a weight of 0. 
We generated data where the base forecasts $\hat{\pmb{Y}}_t$ for those two nodes are extremely far from the truth. 
In the final evaluation, none of the methods will include the performance of those two revised forecasts. 
Incorporating the two highly forecastable nodes into the assessment of accuracy would add large amounts of noise to the results and obscure the performance of the low forecastability nodes.

In Setting 2, we focus on aggregate consistency issues. 
In our simulated setting, we force the `Canada' node to be highly aggregate inconsistent. 
The forecasts of its children are far from the truth and do not add up to a reasonable estimate.
The `Canada' forecast, on the other hand, is very accurate. 
We place a weight of 1 on `Canada', and analyze how the methods create aggregate consistent forecast by learning which nodes to borrow information from. 



\subsection{Simulation Setting 1} \label{sim1}
The set of weights to assess overall accuracy in Setting 1 are $(\frac{1}{6},\frac{1}{6},\frac{1}{6},\frac{1}{6},\frac{1}{6},\frac{1}{6},0,0)$.
The results are displayed in Table \ref{tab:sim1}.
When the reconciliation metric of a dataset is ``easy'', the methods need to only slightly modify the base forecasts $\hat{\pmb{Y}}_t$ in creating the revised forecasts.
This leads to very similar revised forecasts for all the methods. 
Therefore all five methods perform very similarly in the first two rows of the results.

However, when the reconciliation is ``hard'', the performance is driven by how much the methods adjust each of the base forecasts in creating their reconciled forecasts.
In this setting, BR and WLS have the highest accuracy, with the WLS method doing slightly better than BR. 
Unlike the other methods, BR and WLS use a metric of historic or future accuracy to determine which nodes to place a higher weight on resulting in the best performance.
There is not a large difference in using the diagonal $\pmb{Q1}$ or the block diagonal $\pmb{Q2}$ in our method.
The worse performance of BU relative to all other methods is due to the high forecastability of the two children of `Canada'. The forecasts in these two nodes have extremely high aggregate consistency and therefore do not add up to the `Canada' forecast. 
Although these two nodes are not taken into account in the final performance metric, their effect is propagated up the tree. Since OLS weighs every node equally, again the errors in two `Canada' nodes affects the accuracy of the higher levels.

\begin{table}[ht]
\centering
\begin{tabular}{rllrrrrr}
  \hline
 & reconciliation & Q & BR & WLS & BU & TD & OLS \\ 
  \hline
 & easy & Q1 & 1.00 & 1.00 & 1.54 & 0.94 & 1.05 \\ 
   & easy & Q2 & 1.00 & 0.99 & 1.63 & 0.94 & 1.06 \\ \hline
   & hard & Q1 & 1.00 & 0.95 & 479.87 & 1.27 & 23.04 \\ 
   & hard & Q2 & 1.00 & 0.92 & 461.56 & 1.22 & 22.14 \\ 
   \hline
\end{tabular}
\caption{Results for simulation setting 1. We provide the relative error of each competitor method compared to the BR method.}
\label{tab:sim1}
\end{table}

\subsection{Simulation Setting 2} \label{sim2}
To assess accuracy we focus on the `Canada' node by using the weights $(0,0,1,0,0,0,0,0)$. 
`Canada' exemplifies the most interesting dynamics and so we focus on evaluating its performance. 
`Canada' is easy to forecast while its children are not.
In the same manner as in Section \ref{sim1}, we split the data in terms of how ``hard'' or ``easy'' it is to reconcile, and the results are shown in Table \ref{tab:sim2}. 

When the reconciliation metric is ``easy'', all five methods perform similarly.
TD performs slightly better than BR and WLS, however this would depend on how we allocate proportions from the top downwards. In the ``easy'' setting, the base forecasts are initially close to aggregate consistent and so the revised forecasts do not deviate from their initial value. This results in similar revised forecasts and hence similar accuracy across the different methods.

When the reconciliation metric is ``hard'', we see large gains in using our method. 
This is the case with both $\pmb{Q1}$ and $\pmb{Q2}$. The largest gains occur when using the block diagonal matrix $\pmb{Q2}$. This happens because we share information across the nodes in the tree. The $\pmb{Q2}$ matrix allows for co-movement between parent and child nodes. The closer two nodes are the more they can influence each other. This covariance matrix allows for changes of a particular revised forecast to affect other revised forecasts that are closest to it in the hierarchy. 

The gain in performance in BR over WLS is attributed primarily to two factors. The structure of the covariance matrix as discussed above takes into account the correlation between nodes that are close to each other. BR also takes full advantage of the uncertainty in the posterior distributions when constructing the revised forecasts.

The BU method performs poorly because the two children of `Canada' have high forecastability that propagates up the tree. A similar issue occurs with OLS since every node is weighted equally to revise the forecasts. The TD method performs better than OLS and BU in this case because of how it is constructed: the forecasts in the children of `Canada' are not used to revise the `Canada' node which contain noisy base forecasts.

\begin{table}[ht]
\centering
\begin{tabular}{rllrrrrr}
  \hline
 & reconciliation & Q & BR & WLS & BU & TD & OLS \\ 
  \hline
 & easy & Q1 & 1.00 & 1.00 & 1.57 & 0.92 & 1.09 \\ 
   & easy & Q2 & 1.00 & 1.00 & 1.66 & 0.93 & 1.10 \\ \hline
   & hard & Q1 & 1.00 & 1.20 & 18689.04 & 3.48 & 1090.43 \\ 
   & hard & Q2 & 1.00 & 1.38 & 21640.18 & 4.01 & 1262.50 \\ 
   \hline
\end{tabular}
\caption{Results for simulation setting 2. We provide the relative error of each competitor method compared to the BR method.}
\label{tab:sim2}
\end{table}

\subsection{Benefits of Our Method}

We briefly elaborate on two benefits of our method that pertain to the correlation structure and variability of the estimates. 
There is a key difference between the covariance matrices in BR and WLS.
In BR, different methods are used at each historic point in time in order to obtain forecasts sequentially. 
Their performances are then averaged to create the measure of historical accuracy. 
This differs from the creation of the WLS covariance matrix.
The covariance matrix in WLS depends on a single forecast which in turn means it depends on one method. 
In addition, the entries in the WLS covariance matrix are the standard errors of the forecasts which are not measures of accuracy (e.g. differences between the truth and the forecast), but rather measurement errors of the model used.
It is important to note that the WLS algorithm is flexible and allows for different forms of the covariance matrix. The comparison assumes WLS with the default settings. All the above analyses assume a default covariance matrix and will vary depending on what the user chooses to use in WLS.

A second improvement to the covariance matrix is the extension from a diagonal structure.
The block diagonal structure imposed is important because it incorporates similar behavior into its updating scheme within parent/child subtrees.
In other words, if two nodes had similar behavior in the past, the method will assume that they will have similar behavior in the future.
If one node has a narrow posterior distribution and high historical accuracy and a second, correlated node is difficult to forecast, the method will borrow information from the more accurate node to update the more difficult node to forecast. 
As we saw in the simulated results, this approach added an additional improvement of performance for our method over the other methods in the ``hard'' setting.

\begin{figure}[h]
\begin{center}
\includegraphics[width=5.5in]{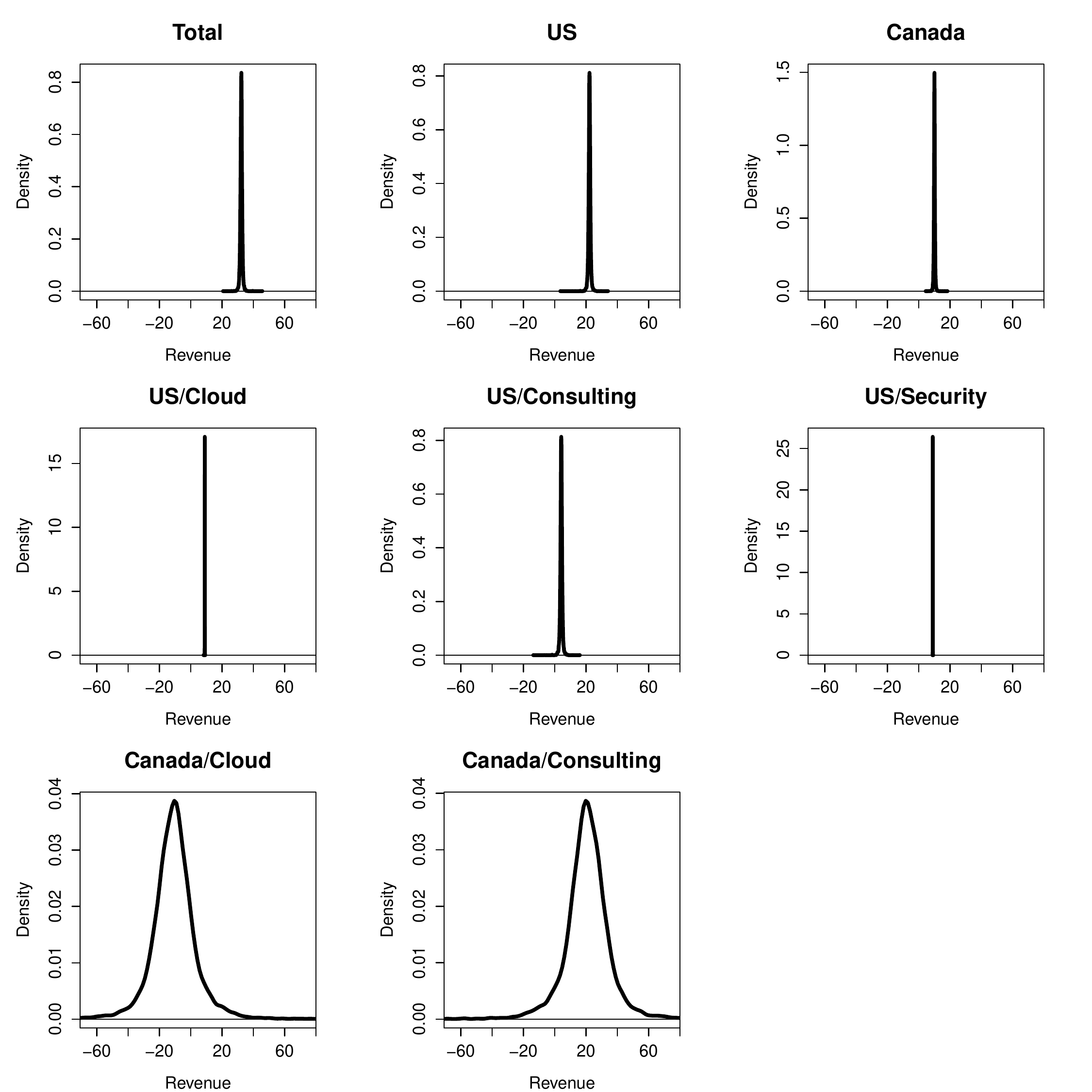}
\end{center}
\caption{Posterior marginal densities for each of the nodes of the hierarchy for one of the simulated datasets for the tree in Figure \ref{fig:hierarchy2}. They reflect the uncertainty for each of the individual nodes of the hierarchy.  
\label{fig:posteriors}}
\end{figure}

Another gain of BR is having full posterior distributions for each of the nodes of the hierarchy.
Figure \ref{fig:posteriors} displays the posterior intervals for all the nodes in our simulated model.
In this example, nodes `Canada/Cloud' and `Canada/Consulting' were extremely noisy and uninformative.
When one compares the posterior intervals one observes large variability in these nodes as compared to all the remaining nodes in the hierarchy. 
This would indicate to the practitioner that there is little information in the updated forecasts for these two nodes, and perhaps will either weigh them less when assessing accuracy across the hierarchy or simply not use them to revise the forecasts. 
Either way, this is an additional source of information beyond the updated forecasts that is highly useful for the user of the model. If the resulting posterior distributions are unacceptably broad for the questions one is trying to answer with the forecasts then this is an indication that additional information must be provided for these nodes, perhaps incorporating other sources of data beyond those included in the problem.    

\FloatBarrier

\section{Revenue Forecasting Application} \label{realdata}

We apply our proposed Bayesian hierarchical method to forecast IBM monthly revenue. The data consists of historical monthly revenue actuals of different products for different markets.
The market dimension will focus on Europe where IBM further divides the continent into eight distinct sub-groups.  These groups are one level up from the country level and the groupings are created to reflect business practices.  
More specifically, countries within these groups have similar growth opportunities in terms of the product offerings.  Other important factors contribute to these groupings, such as similar client demands, language and business culture.

The product dimension will focus on a large services segment of IBM called the Global Technology Services Group (GTS), which handles infrastructure services.
In 2012 revenue from GTS was about 40 billion dollars which accounts for approximately half of all services revenue. 
Services offered by GTS include mobile and network services, cloud and systems services, technology support services and outsourcing services.
These offerings can be divided into three subdivisions (which we cannot disclose due to confidentiality), and this is how we group the data. 
The data is monthly revenue data which cover years 2015 through June of 2017.

We want the setup to be comparable to the simulated data section. 
Since we could not generate different datasets, we changed the length of our dataset to make an ``easy'' and ``hard'' setup (similar to the simulated section).
In the ``easy'' setup, we used 2015-2017 data and had 30 monthly data points, providing more information for use for the base forecasts.
In the ``hard'' setup, we only used 2016-2017 data and had only 18 total datapoints from which to create the base forecasts. 
This results in higher forecastablity and higher aggregate consistency with respect to the ``easy'' setting. 
More specifically, the forecastability was 340 in the shorter dataset and 234 in the longer dataset and the aggregate consistency was 66 in the shorter dataset and 34 in the longer dataset. Forecasts were obtained by first finding the best ARIMA model for the two datasets using the default settings of the auto.arima function of the forecast package in R\citep[][]{forecastR, khandakar2008automatic}.

To obtain a global measure of performance in the IBM hierarchy, we place an equal weight on all of the nodes and calculate the average forecast error. The results are shown in Table \ref{tab:real}.
When the reconciliation setting is ``easy'', OLS, WLS, and BR all perform similarly well.
TD gave a higher performance error; allocation based on historic proportions did not work well in this scenario. 
BU performed the worst out of the five methods; adding up the lowest level nodes propagated the base forecast error in a certain node up to all the levels.
When the reconciliation setting is ``hard'' (i.e., the length of the dataset was short), our method performed best and WLS performed second best.
OLS, TD, and BU all performed significantly worse.
As we saw in the simulated results, when the base forecasts have high aggregate consistency and high forecastability, it is important to take into account the uncertainty as well as historic accuracy of each of the individual nodes in the reconciliation scheme to achieve optimal results. 
Finally, when we use $\pmb{Q2}$, we perform even slightly better than when we use $\pmb{Q1}$, as we are taking advantage of the correlation structure that exists across the different nodes. 

\begin{table}[ht]
\centering
\begin{tabular}{rrllrrrrr}
  \hline
 datapoints & reconciliation & Q & BR & WLS & BU & TD & OLS \\ 
  \hline
 26 & easy & Q1 & 1.00 & 1.01 &1.19  &1.14 & 0.98& \\ 
 26  & easy & Q2 & 1.00 & 0.98 & 1.16 & 1.11 &0.96 & \\ \hline
14   & hard & Q1 & 1.00 & 1.16 & 1.97 & 1.39 & 1.30 & \\ 
 14  & hard & Q2 & 1.00 & 1.17 & 2.00 & 1.41 & 1.32 & \\ 
   \hline
\end{tabular}
\caption{Results for IBM monthly revenue forecasts. We provide the relative error of each competitor method compared to the BR method.}
\label{tab:real}
\end{table}

\section{Conclusion} \label{conc}
Forecast aggregation is a challenge that arises in many contexts, from forecasting revenue \citep{shan2005dynamic, weatherford2003comparison} to estimating economic metrics \citep{hubrich2005forecasting, capistran2010multi} to estimating electricity demand \citep{taieb2016forecasting, fan2008price}.
Regardless of the setting, a practitioner whose data is naturally hierarchically structured, especially when data sources are different for each node, will find themselves in a situation where the numbers they produce will have to be aggregate consistent (whereas their initial forecasts are not). 

To address this issue, we build upon preexisting methods in the literature, specifically \citet{hyndman2011}, to create aggregate consistent forecasts by using a novel Bayesian approach. 
We incorporate historical accuracy and prior information into our Bayesian updating scheme, and additionally consolidate forecasts via a heterogeneous loss function chosen specifically for the problem at hand. 
We also show that by allowing for a block diagonal correlation structure we can improve our updated forecasts, since the past relationships across the nodes influence the updating process. 
Our method accounts for a large variety of different information for creating the final updates. 

We performed a synthetic data analysis and also applied our method to real IBM monthly revenue data for Global Technology Services in Europe.
GTS was split into the three IBM subdivisions and Europe was split into eight official IBM market groupings.
In both cases we compared our method to four common practices: top-down, bottom-up, the ordinary least squares approach of \citet{hyndman2011}, and the weighted least squares approach of \citet{hyndman2016}. 
Both the synthetic and real data results were consistent.
Our method would do similarly to the weighted least squares when the setting is characterized as ``easy'' (easy to forecast and close to aggregate consistent).
However when reconciliation is ``hard'' (which means that forecasting the nodes is more difficult and the original forecasts are not aggregate consistent), our method performs the best (especially when we allow for a block diagonal correlation structure as opposed to a diagonal one).

In addition to performance, there are a number of practical benefits of our method (in addition to aggregate consistent point estimates).
Unlike the alternative approaches, our method provides posterior intervals which account for the uncertainty of the forecasts for each of the nodes.
The practitioner can use these to understand which nodes to contain more accurate information and which require more cautious treatment.  
The heterogeneous loss function is a unique feature in our Bayesian reconciliation scheme.
The practitioner can penalize each of the nodes differently based on which level is most important in their specific context.
Finally, our method does not necessarily assume each of the nodes to be reconciled is independent.
By allowing for a block diagonal correlation structure, the child/parent nodes of the subtree can influence each other's updated forecasts-- especially if they are highly correlated.
We demonstrate in both the simulation and real data sections that this further improves our accuracy over using our method with simply a diagonal covariance matrix. 


One future direction for this work is to apply the Bayesian reconciliation method to other datasets.
These include economic metric forecasting (across different regions and states), energy demand forecasting (also across different locations) are all application areas that could benefit from such an approach. 
A natural extension of this work would be using a similar type of Bayesian method to simultaneously combine multiple forecasts for each of the nodes while also maintaining aggregate consistency. 
It would be an extension of the logic used in this method, however instead of comparing the certainty of the forecasts for each of the nodes, we would have an additional layer where we also compare the forecasts within a specific node.
This extension would also be of importance because often there are multiple competing methods (sometimes unequal in number) for any single node in a hierarchy. 


\section*{Acknowledgements}
This research was supported by the International Institute of Forecasters and SAS Grant to Support Research on Principles of Forecasting.


\section*{References}

\bibliography{ijftemplate}

\end{document}